\documentstyle[aps,psfig,multicol]{revtex}
\newcommand{\bcols}{\ifpreprintsty\else\begin{multicols}{2}\fi}
\newcommand{\ecols}{\ifpreprintsty\else\end{multicols}\fi}
\begin{document}
\draft
\title{Counting statistics of photons produced by electronic shot noise}
\author{C. W. J. Beenakker and H. Schomerus}
\address{Instituut-Lorentz, Universiteit Leiden,
P.O. Box 9506, 2300 RA Leiden, The Netherlands}
\date{August 2000}
\maketitle
\begin{abstract}
A theory is presented for the photodetection statistics of radiation produced
by current fluctuations in a phase-coherent conductor. Deviations are found
from the Poisson statistics that would result from a classical current. For
detection in a narrow frequency interval $\delta\omega$, the photocount
distribution has the negative-binomial form of black-body radiation if
$e\delta\omega$ is less than the mean current $\bar{I}$ in the conductor. When
electronic localization sets in, $\bar{I}$ drops below $e\delta\omega$ and a
different type of super-Poissonian photon statistics results.
\end{abstract}
\bcols

Some interesting recent developments in mesoscopic physics have arisen from the
interplay with quantum optics \cite{Yam99}. To mention two examples, the
Hanbury-Brown--Twiss effect for photons is inspiring the search for its
electronic counterpart \cite{But99}, while single-electron tunneling has been
used to create a single-photon turnstile device \cite{Kim99}. An appealing
subject of research, in line with these developments, is the study of a
mesoscopic conductor through the quantum optical properties of the radiation
produced by the current fluctuations. It is a textbook result \cite{Man95}, due
to Glauber \cite{Gla63b}, that a classical current produces photons with
Poisson statistics. What is the photon statistics for a fully phase-coherent
conductor? That is the fundamental question addressed in this paper.

It is a timely question in view of a recent proposal by Aguado and Kouwenhoven
\cite{Agu00} to use photon-assisted tunneling in a device containing two
quantum dots in series as a detector for the microwave radiation emitted by a
nearby mesoscopic conductor (e.g.\ a quantum point contact). One such device by
itself can only give information on the mean rate of photon production,
calculated in Refs.\ \cite{Agu00,Les97}, but a pair of devices could measure
the time-dependent correlations and hence could detect deviations from Poisson
statistics due to photon bunching \cite{deadtime}.

We will calculate these fluctuations for an idealized model of a photodetector,
the same model that leads to the Glauber formula of photodetection theory
\cite{Gla63a}. In this formula the photocount distribution is expressed as an
expectation value of normally ordered photon creation and annihilation
operators. (Normal ordering means that all creation operators are brought to
the left of the annihilation operators.) We will see that the ordering
inherited by the electron current operators involves not only a normal
ordering, but in addition an ordering of the incoming current with respect to
the outgoing current.

We present a general formula for the variance of the photocount in terms of the
transmission and reflection matrices of the conductor. A particularly simple
result is obtained in the limit that the frequency interval $\delta\omega$ of
the detected radiation is small compared to the mean (particle) current
$\bar{I}/e$ through the conductor: The photocount distribution $P(n)$ for a
long counting time $\tau$ is then proportional to the negative-binomial
coefficient ${n+\nu-1\choose n}$ (with $\nu=\tau\delta\omega/2\pi\gg 1$)
\cite{negativebinomial}. This is the photonic counterpart of the (positive)
binomial counting distribution for electrical charge
\cite{Lev93,Muz94,Naz99,And00}. In the localized regime the condition
$\delta\omega\ll\bar{I}/e$ breaks down and a different non-Poissonian
distribution results.

Starting point of our analysis is an expression for the photocount distribution
as a time-ordered expectation value of the electric field operator
\cite{Gla63a},
\begin{eqnarray}
&&P(n)=\langle{\cal T}_{\pm}\frac{1}{n!}W^{n}{\rm
e}^{-W}\rangle,\label{PnWdef}\\
&&W=\int_{0}^{\infty}d\omega\,\alpha(\omega)
\int\!\!\!\!\int_{0}^{\tau}dt_{-}dt_{+}\,{\rm e}^{i\omega(t_{+}-t_{-})}
E(t_{-})E(t_{+}).\nonumber
\end{eqnarray}
Here $P(n)$ is the probability to detect $n$ photons in a time interval $\tau$
and $E(t)$ is the operator of the detected mode of the electric field in the
Heisenberg picture. (We assume for simplicity that a single mode is detected.)
The detector has sensitivity $\alpha(\omega)$ at frequency $\omega>0$. The
symbol ${\cal T}_{\pm}$ indicates the Keldysh time-ordering of the
time-dependent operators: times $t_{-}$ to the left of times $t_{+}$, earlier
$t_{-}$ to the left of later $t_{-}$, earlier $t_{+}$ to the right of later
$t_{+}$.

The Glauber formula mentioned in the introduction is obtained from Eq.\
(\ref{PnWdef}) by substituting the free-field expression
\begin{equation}
E_{\rm free}(t)\propto\int_{0}^{\infty}d\omega\, \left(a^{\dagger}(\omega){\rm
e}^{i\omega t}+a(\omega){\rm e}^{-i\omega t}\right)\label{Efree}
\end{equation}
for $E(t)$ and by making the rotating-wave approximation [neglecting ${\rm
e}^{i(\omega+\omega')t}$, retaining ${\rm e}^{i(\omega-\omega')t}$]. The time
ordering of the electric field then becomes normal ordering of the photon
operators $a^{\dagger},a$. Our goal is instead to go from Eq.\ (\ref{PnWdef})
to an expression in terms of the electron operators $c^{\dagger},c$ that
constitute the current operator $I(t)$.

The electron and photon degrees of freedom are coupled in the Hamiltonian via a
term $-\int d{\bf r}\,{\bf j}({\bf r},t)\cdot{\bf A}({\bf r},t)$, where $\bf j$
is the electron current density operator and $\bf A$ the electromagnetic vector
potential. This bilinear coupling leads to a linear integral relation between
$E$ and $I$,
\begin{equation}
E(t)=E_{\rm
free}(t)+\int_{-\infty}^{\infty}dt'\,g(t-t')I(t').\label{EIrelation}
\end{equation}
The propagator $g(t)$ vanishes for $t<0$ because of causality. (We are
neglecting retardation of the electromagnetic radiation.) We assume that
electrons and photons are uncoupled for $t\rightarrow -\infty$, the photons
starting out in the vacuum state. Substitution of Eq.\ (\ref{EIrelation}) into
Eq.\ (\ref{PnWdef}) leads to a correlator involving the non-commuting operators
$E_{\rm free}$ and $I$. Fleischhauer has shown \cite{Fle98} that the vacuum
term $E_{\rm free}$ may be removed from the correlator if the Keldysh
time-ordering of $E$ is carried over to $I$:
\begin{eqnarray}
W&=&\int_{0}^{\infty}d\omega\,\alpha(\omega)
\int\!\!\!\!\int_{0}^{\tau}dt'dt''\,{\rm
e}^{i\omega(t''-t')}\int\!\!\!\!\int_{-\infty}^{\infty} dt_{-}dt_{+}\nonumber\\
&&\mbox{}\times g(t'-t_{-})g(t''-t_{+})I(t_{-})I(t_{+}). \label{WIrelation}
\end{eqnarray}

\begin{figure}
\centerline{\psfig{figure=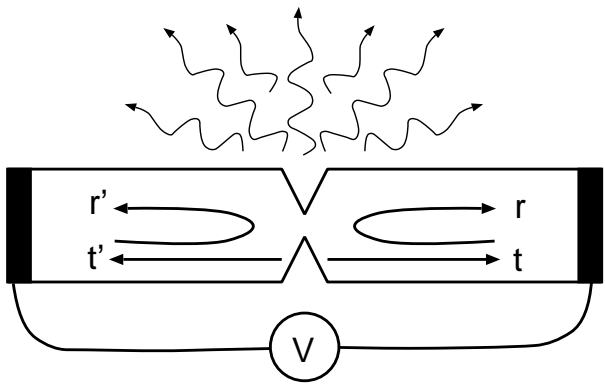,width=8cm}\medskip}
\caption[]
{Illustration of the scattering geometry studied in the text. An electrical
current flowing through a constriction emits microwave radiation that is
absorbed by a nearby detector.
\label{ecountdiagram}}
\end{figure}

To leading order in $g$ we may neglect the coupling to the photons in the time
dependence of $I(t)$. For free electrons the time dependence is given by
\cite{But90}
\begin{eqnarray}
&&I(t)=\frac{1}{2\pi}\int\!\!\!\!\int d\varepsilon d\varepsilon'\,{\rm
e}^{i(\varepsilon-\varepsilon')t}
c^{\dagger}(\varepsilon)M(\varepsilon,\varepsilon')c(\varepsilon'),
\label{Icrelation}\\
&&M(\varepsilon,\varepsilon')=S^{\dagger}(\varepsilon)DS(\varepsilon')-D,
\label{Mdef}\\
&&D=\left(\begin{array}{cc}
0&0\\0&1\end{array}\right),\;\;
S=\left(\begin{array}{cc}
r'&t'\\t&r\end{array}\right),\label{DSdef}
\end{eqnarray}
using units such that $\hbar=1$ and $e=1$. We have introduced the scattering
matrix $S(\varepsilon)$ (with $N\times N$ reflection and transmission
submatrices $r,r',t,t'$) of the $N$ propagating modes at energy $\varepsilon$
(relative to the Fermi energy at $\varepsilon=0$). The scattering geometry is
illustrated in Fig.\ \ref{ecountdiagram}. The detection matrix $D$ selects the
current in one of the two leads, arbitrarily chosen to be the right lead in
Eq.\ (\ref{DSdef}). The total current $I=I_{\rm out}-I_{\rm in}$ is then the
difference of the current $I_{\rm out}$ coming from the left and the current
$I_{\rm in}$ coming from the right. These two currents $I_{\rm out}$ and
$I_{\rm in}$ are defined as in Eq.\ (\ref{Icrelation}), with the matrix $M$
replaced by $S^{\dagger}DS$ and $D$, respectively.

The separation of $I$ into outgoing and incoming current operators is
convenient because they have simple commutation relations: (i) $I_{\rm out}(t)$
commutes with $I_{\rm out}(t')$; (ii) $I_{\rm in}(t)$ commutes with $I_{\rm
in}(t')$; (iii) $I_{\rm out}(t)$ commutes with $I_{\rm in}(t')$ if $t<t'$. It
follows that Keldysh time-ordering of the current operators is the same as an
ordering whereby the operators $I_{\rm in}(t_{-})$ are moved to the left and
$I_{\rm in}(t_{+})$ to the right of all other operators --- irrespective of the
values of the time arguments.

Now that we have liberated ourselves from the time ordering we are free to take
Fourier transforms,
\begin{eqnarray}
&&P(n)=\langle{\cal O}\frac{1}{n!}W^{n}{\rm e}^{-W}\rangle,\;\;
W=\int_{0}^{\infty}d\omega\,\alpha(\omega)
U^{\dagger}(\omega)U(\omega),\nonumber\\
&&U(\omega)=\int_{-\infty}^{\infty}\frac{d\omega'}{2\pi}\,
K(\omega-\omega')g(\omega') [I_{\rm out}(\omega')-I_{\rm in}(\omega')].\label{Udef}
\end{eqnarray}
The Fourier transforms of $g,I_{\rm out},I_{\rm in}$ are defined as
$f(\omega)=\int_{-\infty}^{\infty}dt\,{\rm e}^{i\omega t}f(t)$, and we have
abbreviated $K(\omega)=\int_{0}^{\tau}dt'\,{\rm e}^{i\omega t'}$. The symbol
$\cal O$ indicates ordering of the current operators from left to right in the
order $I_{\rm in}^{\dagger},I_{\rm out}^{\dagger},I_{\rm
out}^{\vphantom\dagger},I_{\rm in}^{\vphantom\dagger}$. According to Eq.\
(\ref{Icrelation}), the current operators are related to the electron creation
and annihilation operators by
\begin{mathletters}
\label{Icrelation2}
\begin{eqnarray}
I_{\rm out}(\omega)&=&\int d\varepsilon\,
c^{\dagger}(\varepsilon)S^{\dagger}(\varepsilon) DS(\varepsilon+\omega)
c(\varepsilon+\omega),\label{Icrelation2a}\\
I_{\rm in}(\omega)&=&\int d\varepsilon\,
c^{\dagger}(\varepsilon)Dc(\varepsilon+\omega).\label{Icrelation2b}
\end{eqnarray}
\end{mathletters}%
Eqs.\ (\ref{Udef}) and (\ref{Icrelation2}) form the required relation between
the photocount distribution and the electron creation and annihilation
operators.

The mean photocount $\bar{n}$ has been studied before, by Lesovik and Loosen
\cite{Les97} and by Aguado and Kouwenhoven \cite{Agu00}. To make contact with
that work we take the experimentally relevant limit of a long detection time
$\tau$. We may then discretize the frequencies as $\omega_{n}=n\times
2\pi/\tau$. In this discretization the kernel $K$ becomes a Kronecker delta,
$K(\omega_{n}-\omega_{m})=\tau\delta_{nm}$, hence
$U(\omega_{n})=g(\omega_{n})I(\omega_{n})$. The factorial moments
\begin{equation}
\langle n^{p}\rangle_{\rm
f}\equiv\overline{n(n-1)(n-2)\cdots(n-p+1)}=\langle{\cal
O}W^{p}\rangle\label{nfactdef}
\end{equation}
of the distribution (\ref{Udef}) in the long-time limit take the form
\cite{note1}
\begin{equation}
\langle n^{p}\rangle_{\rm f}=\langle{\cal
O}[{\textstyle\int}d\omega\,\gamma(\omega)
I^{\dagger}(\omega)I(\omega)]^{p}\rangle,\label{nfactresult}
\end{equation}
with $\gamma(\omega)=\alpha(\omega)|g(\omega)|^{2}$. For the first moment the
ordering operator ${\cal O}$ can be omitted and we find
\begin{eqnarray}
&&\bar{n}=\int_{0}^{\infty}d\omega\,\gamma(\omega)\langle
I^{\dagger}(\omega)I(\omega)\rangle,\label{nbarresulta}\\
&&\langle I^{\dagger}(\omega)I(\omega)\rangle=
\tau\int_{-\infty}^{\infty}dt\,{\rm e}^{i\omega t}\langle
I(0)I(t)\rangle,\label{nbarresultb}
\end{eqnarray}
in agreement with Refs.\ \cite{Agu00,Les97}.

For the double-quantum-dot photodetector of Aguado and Kouwenhoven \cite{Agu00}
the response function $\gamma(\omega)$ is sharply peaked at the frequency
$\Omega$ of the inelastic tunneling transition, with a width
$\delta\omega\ll\Omega$. Its integrated magnitude
$\gamma(\Omega)\delta\omega\simeq(Z\Gamma/\Omega)^{2}$ depends on the impedance
$Z$ (in units of $h/e^{2}$) of the inductive coupling and the ratio of $\Omega$
and the tunnel rate $\Gamma$ between the quantum dots. (Typically, $Z\lesssim
1$ and $\Gamma\ll\Omega$, so that $\gamma\delta\omega\ll 1$.) In that device
the inelastic transition can be either upwards or downwards in energy,
corresponding, respectively, to the absorption or emission of a photon. Here we
consider only the case of photodetection by absorption, which is the relevant
case for the study of shot noise in the conductor \cite{note2}.

We now go beyond the first moment to study the entire photocount distribution.
We note that $P(n)$ in Eq.\ (\ref{Udef}) would be simply a Poisson
distribution,
\begin{equation}
P_{\rm Poisson}(n)=\frac{1}{n!}\bar{n}^{n}{\rm e}^{-\bar{n}}, \label{PPoisson}
\end{equation}
if the current $I$ would be a classical quantity instead of a quantum
mechanical operator. This is in accordance with Glauber's finding \cite{Gla63b}
that the radiation produced by a classical current is in a coherent state
(since a coherent state has Poisson statistics).

To find the deviations from the Poisson distribution due to quantum statistics,
let us consider the case of a conductor connecting two electron reservoirs in
thermal equilibrium at temperature $T$. The system is brought out of
equilibrium by application of a voltage difference $V$ between the left and
right reservoirs. Expectation values are given by the Fermi function
$f(\varepsilon)=(e^{\varepsilon/T}+1)^{-1}$,
\begin{equation}
\langle
c^{\dagger}_{i}(\varepsilon)c^{\vphantom\dagger}_{j}(\varepsilon')\rangle=
\delta_{ij} \delta(\varepsilon-\varepsilon')f(\varepsilon-\mu_{i}),
\label{thermalaverage}
\end{equation}
with higher order expectation values obtained by pairwise averaging. The
potential $\mu_{i}$ equals $V$ for the left reservoir (mode indices
$i=1,2,\ldots ,N$) and 0 for the right reservoir ($i=N+1,N+2,\ldots ,2N$).

For simplicity we restrict ourselves to zero temperature, when
$f(\varepsilon-\mu_{i})$ becomes the step function
$\theta(\mu_{i}-\varepsilon)$ (equal to 1 for $\varepsilon<\mu_{i}$ and 0 for
$\varepsilon>\mu_{i}$). The mean and variance of the photocount are then given
by
\begin{eqnarray}
&&\bar{n}=\frac{\tau}{2\pi}\int_{0}^{V} d\omega\,{\cal N}, \label{barnresult}\\
&&{\rm Var}\,n=\bar{n}+\frac{\tau}{2\pi}\int_{0}^{V} d\omega\,{\cal N}^{2}
+\frac{\tau}{2\pi}\int\!\!\!\!\int_{0}^{V} d\omega d\omega'\,
\gamma(\omega)\gamma(\omega') \nonumber\\
&&\;\;\;\;\;\;\;\;\;\;\;\;\;\mbox{}\times\int_{0}^{V} d\varepsilon\,{\rm
Tr}\,(A_{1}-A_{2}-A_{3}),\label{varnresult}\\
&&{\cal N}=\gamma(\omega)\int_{\omega}^{V} d\varepsilon\,{\rm
Tr}\,\tau_{\varepsilon}\rho_{\varepsilon-\omega},\\
&&A_{1}=\tau_{\varepsilon}(1-\tau_{\varepsilon-\omega}-
\tau_{\varepsilon-\omega'}) \rho_{\varepsilon-\omega-\omega'} 
(1-\tau_{\varepsilon-\omega}-\tau_{\varepsilon-\omega'})\nonumber\\
&&\;\;\;\;\;\;\;\;\;\;\;\;\;\mbox{}\times\theta(\varepsilon-\omega-\omega'),\\
&&A_{2}=\tau_{\varepsilon}\rho_{\varepsilon-\omega}
\tau_{\varepsilon}\rho_{\varepsilon-\omega'}
\theta(\varepsilon-\omega)\theta(\varepsilon-\omega'),\\
&&A_{3}=\rho_{\varepsilon}\tau_{\varepsilon+\omega} \rho_{\varepsilon}
\tau_{\varepsilon+\omega'}\theta(V-\varepsilon-\omega)
\theta(V-\varepsilon-\omega').
\end{eqnarray}
We have abbreviated
$\tau_{\varepsilon}=t(\varepsilon)t^{\dagger}(\varepsilon)=
1-\rho_{\varepsilon}$.

The formula (\ref{barnresult}) for the mean photocount is known
\cite{Agu00,Les97}, the result (\ref{varnresult}) for the variance is new. The
first term $\bar{n}$ on the right-hand-side corresponds to Poisson statistics.
The other terms describe the excess noise, consisting of one term containing a
single integral over frequency and three more terms containing double frequency
integrals. For narrow-band detection the single frequency integral dominates.
More precisely, if $\gamma(\omega)$ is non-zero in a narrow frequency range
$\delta\omega\ll V$, then
\begin{equation}
{\rm Var}\,n=(\tau\delta\omega/2\pi){\cal N}(1+{\cal N}).\label{varnnarrowband}
\end{equation}
The correction terms from the double frequency integrals are smaller by a
factor $\delta\omega/\bar{I}$, with $\bar{I}\propto V\,{\rm Tr}\,tt^{\dagger}$
the mean electrical current flowing between the reservoirs.

In this regime of narrow-band detection one can also calculate easily the
higher order moments of the photocount. The factorial cumulants
$\langle\!\langle n^{p}\rangle\!\rangle_{\rm f}$ are given by
\begin{equation}
\langle\!\langle n^{p}\rangle\!\rangle_{\rm f}=
(\tau\delta\omega/2\pi)(p-1)!\,{\cal N}^{p}. \label{npnarrowband}
\end{equation}
The probability distribution $P(n)$ can be reconstructed from the factorial
cumulants via the generating function
$F(\xi)=\sum_{p=1}^{\infty}(\xi^{p}/p!)\langle\!\langle
n^{p}\rangle\!\rangle_{\rm f}$, by means of the formula
\begin{equation}
P(n)=\frac{1}{n!}\lim_{\xi\rightarrow -1}\frac{d^{n}}{d\xi^{n}}e^{F(\xi)}.
\label{PFrelation}
\end{equation}
The probability distribution corresponding to Eq.\ (\ref{npnarrowband}) is
\begin{equation}
P(n)={n+\nu-1\choose n}\frac{{\cal N}^{n}}{(1+{\cal
N})^{n+\nu}},\label{Pnegbin}
\end{equation}
which is the negative-binomial distribution with $\nu=\tau\delta\omega/2\pi$
degrees of freedom. (For non-integer $\nu$ the binomial coefficient should be
interpreted as a ratio of Gamma functions.) It approaches the Poisson
distribution (\ref{PPoisson}) in the limit ${\cal N}\rightarrow 0$,
$\nu\rightarrow\infty$, at fixed $\bar{n}=\nu{\cal N}$. The negative-binomial
distribution is known in quantum optics as the distribution of black-body
radiation \cite{Man95}. The role of $\cal N$ is then played by the Bose
function $(e^{\omega/T}-1)^{-1}$ at the temperature of the black body. In both
contexts $\cal N$ is a small parameter, and hence the corrections to Poisson
statistics are small: for a black body, $\cal N$ is small because $\omega\gg
T$, while for the electrical conductor ${\cal N}\simeq \gamma\bar{I}\ll 1$.

We have seen that the negative-binomial distribution results if contributions
of order $\delta\omega/\bar{I}$ can be neglected. If the electrical conductor
is metallic, it is sufficient that $\delta\omega\ll V$, since the conductance
(in units of $e^{2}/h$) is greater than 1 in a metal. In the localized regime,
on the contrary, the conductance becomes exponentially small and terms of order
$\delta\omega/\bar{I}$ start playing a role --- even if $\delta\omega\ll V$. To
illustrate this, let us assume that the transmission through the conductor is
so small that only terms linear in $\tau_{\varepsilon}$ need to be retained.
This leaves only the term $A_{1}$ in Eq.\ (\ref{varnresult}), so that
\begin{equation}
{\rm Var}\,n=\bar{n}+\frac{\tau}{2\pi}(\gamma\delta\omega)^{2}\theta(V-2\omega)
\int_{2\omega}^{V} d\varepsilon\,{\rm Tr}\,\tau_{\varepsilon}.
\label{varnlocalized}
\end{equation}
More generally, the factorial cumulants are given by
\begin{equation}
\langle\!\langle n^{p}\rangle\!\rangle_{\rm
f}=\frac{\tau}{2\pi}(\gamma\delta\omega)^{p}\theta(V-p\omega)
\int_{p\omega}^{V} d\varepsilon\,{\rm
Tr}\,\tau_{\varepsilon}.\label{nplocalized}
\end{equation}
The full distribution $P(n)$ can be reconstructed by means of the inversion
formula (\ref{PFrelation}), but does not have a simple closed-form expression.
We note that the deviations from Poisson statistics are again small because
$\gamma\delta\omega\ll 1$.

Much larger deviations can be obtained if coherent radiation from a reference
source at frequency $\Omega$ is superimposed prior to detection. Such homodyne
detection not only amplifies the deviations from Poisson statistics, it also
provides a way to measure the counting distribution of electrical charge. To
see this, we note that homodyning amounts to the replacement of the current
operator $I(\omega)$ by $I(\omega)+I_{0}\delta(\omega-\Omega)$, where $I_{0}$
is some known classical current (a c-number, not an operator). For
$I_{0}\gg\bar{I}$ we find from Eq.\ (\ref{nfactresult}) for the factorial
cumulants of the photocount distribution the expression
\begin{equation}
\langle\!\langle n^{p}\rangle\!\rangle_{\rm f}= \delta_{p1}(\tau \gamma
I_{0}^{2}/2\pi)+ (2\gamma I_{0})^{p}\langle\!\langle Q^{p}\rangle\!\rangle,
\label{nQrelation}
\end{equation}
where $\langle\!\langle Q^{p}\rangle\!\rangle$ is defined through the
generating function
\[
\sum_{p=1}^{\infty}\frac{\xi^{p}}{p!}\langle\!\langle Q^{p}\rangle\!\rangle=
\ln\left\langle e^{-I_{\rm in}^{\dagger}\xi/2}e^{I_{\rm
out}^{\dagger}\xi/2}e^{I_{\rm out}^{\vphantom\dagger}\xi/2}e^{-I_{\rm
in}^{\vphantom\dagger}\xi/2}\right\rangle.
\]
Comparison with Refs.\ \cite{Lev93,Muz94} shows that $\langle\!\langle
Q^{p}\rangle\!\rangle$ has the interpretation of the cumulant of the charge $Q$
transmitted through the conductor (in units of $e$). Levitov and Lesovik
\cite{Lev93} proposed to measure the charge counting distribution from the
precession of a spin $\frac{1}{2}$ coupled to the current. The photodetection
scheme proposed here provides an alternative, and possibly more practical, way
to count the charge without breaking the circuit.

In summary, we have presented a solution to the classic problem of the
statistics of radiation produced by a fluctuating current. We go beyond the
textbook result by considering a fully phase-coherent conductor, and find small
deviations from the Poisson statistics associated with a classical current
source. The deviations might be measured using an array of double-quantum-dot
photodetectors \cite{Agu00}. The deviations can be amplified by homodyning, in
which case they are directly related to the statistics of the electrical charge
transmitted through the conductor \cite{Lev93}. We have given specific results
for a conductor between normal reservoirs in thermal equilibrium, but our
general formulas can be applied to more special current sources as well. The
applications to entangled \cite{Bur00} or superconducting \cite{Tor99,Jeh00}
electrons seem particularly interesting.

We have benefitted from discussions with L. S. Levitov, E. G. Mishchenko, B. A.
Muzykantskii, and Yu.\ V. Nazarov. This work was supported by the Dutch Science
Foundation NWO/FOM.

\ecols

\begin{references}
\bibitem{deadtime} The double-quantum-dot photodetector has a dead time of
$1/\delta\omega$, set by the width $\hbar\delta\omega$ of the inelastic
transition. Photon bunching occurs on the time scale of the inverse band width
of the detected radiation, which is also $1/\delta\omega$. This is why one
would need at least two such devices to measure deviations from Poisson
statistics due to photon bunching. The theory presented here, appropriate for
an array of devices, does not include dead-time effects.
\bibitem{negativebinomial} This binomial coefficient is called ``negative''
because it appears as the expansion coefficient of $1-x$ to a negative power
$-\nu$. It counts the number of partitions of $n$ bosons among $\nu$ states.
\bibitem{note1} We have reverted to continuous frequency for ease of notation,
but the discretization should be borne in mind when encountering a delta
function at zero frequency argument:
$\delta(\omega-\omega')\rightarrow\tau/2\pi$ when $\omega\rightarrow\omega'$.
This ensures that the factorial cumulants $\langle\!\langle
n^{p}\rangle\!\rangle_{\rm f}$ increase linearly with the detection time $\tau$
in the long-time limit.
\end{references}
\end{document}